\documentclass[10pt]{iopart}
\usepackage{amsfonts}
\usepackage{cite}
\usepackage{graphicx}
\usepackage{subfig}
\usepackage{booktabs}
\usepackage{multirow}
\usepackage{paralist}

\begin{document}
\title
[Cross-Correlation Based Discriminant Criterion for Channel Selection in MI BCI Systems]
{Cross-Correlation Based Discriminant Criterion for Channel Selection in Motor Imagery BCI Systems}

\author{Jianli~Yu and Zhu~Liang~Yu~}

\begin{abstract}
\textit{Objective.}
Many electroencephalogram (EEG)-based brain-computer interface (BCI) systems use a large amount of channels for higher performance, which is time-consuming to set up and inconvenient for practical applications.
Finding an optimal subset of channels without compromising the performance is a necessary and challenging task.
\textit{Approach.}
In this article, we proposed a cross-correlation based discriminant criterion (XCDC) which assesses the importance of a channel for discriminating the mental states of different motor imagery (MI) tasks.
Channels are ranked and selected according to the proposed criterion.
The efficacy of XCDC is evaluated on two motor imagery EEG datasets.
\textit{Main results.}
In both datasets, XCDC significantly reduces the amount of channels without compromising classification accuracy compared to the all-channel setups.
Under the same constraint of accuracy, the proposed method requires fewer channels than existing channel selection methods based on Pearson's correlation coefficient and common spatial pattern.
Visualization of XCDC shows consistent results with neurophysiological principles.
\textit{Significance.}
This work proposes a quantitative criterion for assessing and ranking the importance of EEG channels in MI tasks and provides a practical method for selecting the ranked channels in the calibration phase of MI BCI systems, which alleviates the computational complexity and configuration difficulty in the subsequent steps, leading to real-time and more convenient BCI systems.
\end{abstract}

\maketitle
\ioptwocol

\section{Introduction}\label{sec.introduction}
Brain-computer interface (BCI) systems bridge human intention and computers, enabling people to communicate with external devices through brain activity \cite{vidal1973toward}. 
They have been employed on wheelchair control \cite{tsui2011wheelchair,long2012wheelchair,li2013wheelchair}, limb-assistive robots \cite{meng2016limb,he2018limb}, post-stroke rehabilitation \cite{ang2012rehab,ang2013rehab,tung2013rehab},  video games \cite{tangermann2008pinball,pires2011tetris,van2013wow} and so on. 
Eletroencephalography (EEG) is widely used in BCI systems, because it is a non-invasive and cost-effective method to acquire brain signals from multiple channels with high temporal resolution \cite{padfield2019eeg}. 
A popular type of mental task using EEG-based BCI is motor imagery (MI), which requires users to imagine specific limb or muscle movements without overt actions. 
These imaginations generate oscillatory activities, such as event-related synchronization (ERS) and event-related desynchronization (ERD) in sensorimotor rhythms, which can be recognized by MI-BCI systems \cite{pfurtscheller1999event,pfurtscheller2001motor}.

In practice, most BCI systems collect EEG signals from multiple electrodes on the scalp to achieve higher performance.
However, setting up a BCI system with a large amount of electrodes is cumbersome and time-consuming, which is a major obstacle of introducing BCI into daily applications.
Moreover, numerous EEG channels bring problems like redundant information, noise and high computation complexity \cite{arvaneh2011optimizing, qi2020spatiotemporal}.
Therefore, selecting an appropriate subset of channels is of great importance in the design of BCI systems.
This article focuses on addressing the problem of channel selection in binary classification of motor imagery EEG signals.

In the literature, various channel selection methods have been proposed. 
Some techniques measure the importance of channels based on criteria related to the performance of specific classifiers, these methods can be characterized as wrapping techniques \cite{alotaiby2015review, baig2020filtering}.
Lal \textit{et al.} \cite{lal2004support} determines the importance of a channel based on the influence the channel has on the margin of a trained support vector machine (SVM) classifier, and channels of least importance are removed iteratively using Recursive Feature Elimination (RFE).
Support vector channel selection method is also employed by Zhou and Yedida \cite{zhou2007channel} with Time-frequency Synthesized Spatial Patterns (TFSP) as input.
Sannelli \textit{et al.} \cite{sannelli2010optimal} uses test errors of an Linear Discriminant Analysis (LDA) classifier as the criterion for EEG channel removal and selection.
These methods are highly coupled with the classifiers, which may lead to worse performance if the classifiers perform poorly.

In some techniques, EEG channels are selected or ranked according to independent evaluation criteria without the assistance of classifiers, these methods can be grouped into filtering techniques.
EEG signal statistics, such as mutual information \cite{lan2006salient, wang2019channel} and Fisher's criterion \cite{tam2011minimal} are employed in the literature as the criteria for channel selection.
Jin \textit{et al.} \cite{jin2019correlation} measures similarity between channels using Pearson's correlation coefficient, and emphasizes the channels with highest overall similarity to others. Such methods are prone to common noise across channels.
Barachant and Stephane \cite{barachant2011channel} remove channels iteratively down to a given number based on the Riemannian distance between covariance matrices of EEG signals of different classes.
However, this kind of greedy manner is computationally expensive and susceptible to local minima \cite{qi2020spatiotemporal}.
Recent years, common spatial patterns (CSP) is widely used as a spatial filtering algorithm in BCI systems \cite{chen2020deep, ang2012filter}.
The spatial filter coefficients of CSP are also utilized in \cite{wang2006common, tam2011minimal, feng2019optimized} for EEG channel selection, $l_1$-norm \cite{farquhar2006regularised,yong2008sparse} and $l_1/l_2$-norm \cite{arvaneh2011optimizing,qi2020spatiotemporal} regularization are introduced to obtain sparser CSP filters, which implicitly select channels.

In this paper, we proposed a cross-correlation based discriminant criterion (XCDC), which efficiently provides a quantitative assessment of the ability of an EEG channel to discriminate between motor imagery tasks.
XCDC measures similarity between signals by cross-correlation, and emphasizes the electrodes that: 
\begin{inparaenum}[1)]
    \item signals of the same class show larger similarity, and 
    \item signals of different classes differ more obviously.
\end{inparaenum}
After ranking the channels according to XCDC, signals from the channels with highest discriminant criterion are chosen as the input of a convolutional neural network (CNN) classifier, which further evaluates the credibility of the chosen channels by classification accuracy.
The performance of XCDC is evaluated and compared to CCS \cite{jin2019correlation} and CSP-rank \cite{tam2011performance} on two publicly available motor imagery EEG datasets.

The remainder of this article is constructed as follows.
Section \ref{sec.methodology} describes the proposed method and the classifier used for evaluation. Section \ref{sec.results} presents descriptions of datasets and experiments, as well as experimental results. More insights into XCDC are offered in section \ref{sec.discussions}, and section \ref{sec.conclusion} concludes this article.

\section{Methodology}\label{sec.methodology}
\subsection{Cross-Correlation Based Discriminant Criterion}
In the procedure of motor imagery EEG classification, signals of the same class of MI task should contain similar features and vice versa.
Therefore, we can assess a channel's discriminative ability based on the similarity between signals of multiple classes from that channel.
In light of this hypothesis, we proposed a channel selection method based on cross-correlation between signals.

Given two finite discrete time series $x(i)$ and $y(j)$ $(i\in\{0, 1, \dots, n-1\}, j\in\{0, 1, \dots, m-1\}, m\geq{n})$, their cross-correlation series is given as follows:
\begin{equation}\label{eqn.1}
    r_{xy}(k) = \sum_{i=0}^{n-1}x(i)y(i+k), k=0, 1, \dots, m-n+1
\end{equation}
Consider two EEG signals $x$ and $y$ of length $T$ (i.e. $m=n=T$). 
In the proposed method, zero-padding is performed at both sides of $y$ to ensure that the length of the cross-correlation series is identical to the length of the original signals.
Then the similarity between $x$ and $y$ is measured by the maximum value of their cross-correlation:
\begin{equation}\label{eqn.2}
    S(x, y) = \mathbf{max}(r_{x\hat{y}}(k)),k=0, 1, \dots, T-1
\end{equation}
where $S(x, y)$ gives the similarity between $x$ and $y$, and $\hat{y}$ denotes signal $y$ with zero-padding.
Note that for signals of the same length, the similarity metric given by (\ref{eqn.2}) is symmetric, i.e. $S(x, y)=S(y, x)$.

In order to assess the discriminative ability of an EEG channel, similarities within class and between classes are taken into consideration.
For an $M$-class motor imagery EEG classification problem, given $N$ trials of EEG signals and their class labels, the discriminant score of a channel is derived as follows.
First, z-score normalization is used to avoid the effect of signal amplitudes on cross-correlation:
\begin{equation}\label{eqn.3}
    \tilde{x}_i = \frac{{x_i}-\bar{x}_i}{\sigma_{x_i}}, i=1, 2, \dots, N
\end{equation}
where $x_i$ is the signal of the $i$th trial collected from the channel of interest, $\bar{x}_i$ and $\sigma_{x_i}$ denotes the mean and standard deviation of $x_i$ respectively.
Second, the within-class \textit{similarity} $R_w$ and between-class \textit{dissimilarity} $R_b$ are obtained by averaging:
\begin{equation}\label{eqn.4}
    R_w = \mathbf{mean}({S(\tilde{x}_i, \tilde{x}_j | c_i = c_j)})
\end{equation}
\begin{equation}\label{eqn.5}
    R_b = -\mathbf{mean}({S(\tilde{x}_i, \tilde{x}_j | c_i \neq c_j)})
\end{equation}
where $c_i$ denotes the class label of the $i$th trial.
The discriminant score $D$ is then derived:
\begin{equation}\label{eqn.6}
    D = \lambda R_w + (1 - \lambda) R_b
\end{equation}
in which $\lambda$ is a weighting hyperparameter to be determined in the experiments.

Channels are ranked in a descending order according to their discriminant score $D$ after obtaining $D$ for every channel using (\ref{eqn.6}).
Given a designated number $k$, signals from the top-$k$ channels in the ranking are chosen as the input for subsequent classification.

\subsection{Classification}
\begin{figure}[h]
	\centering
	\includegraphics[width=0.95\linewidth]{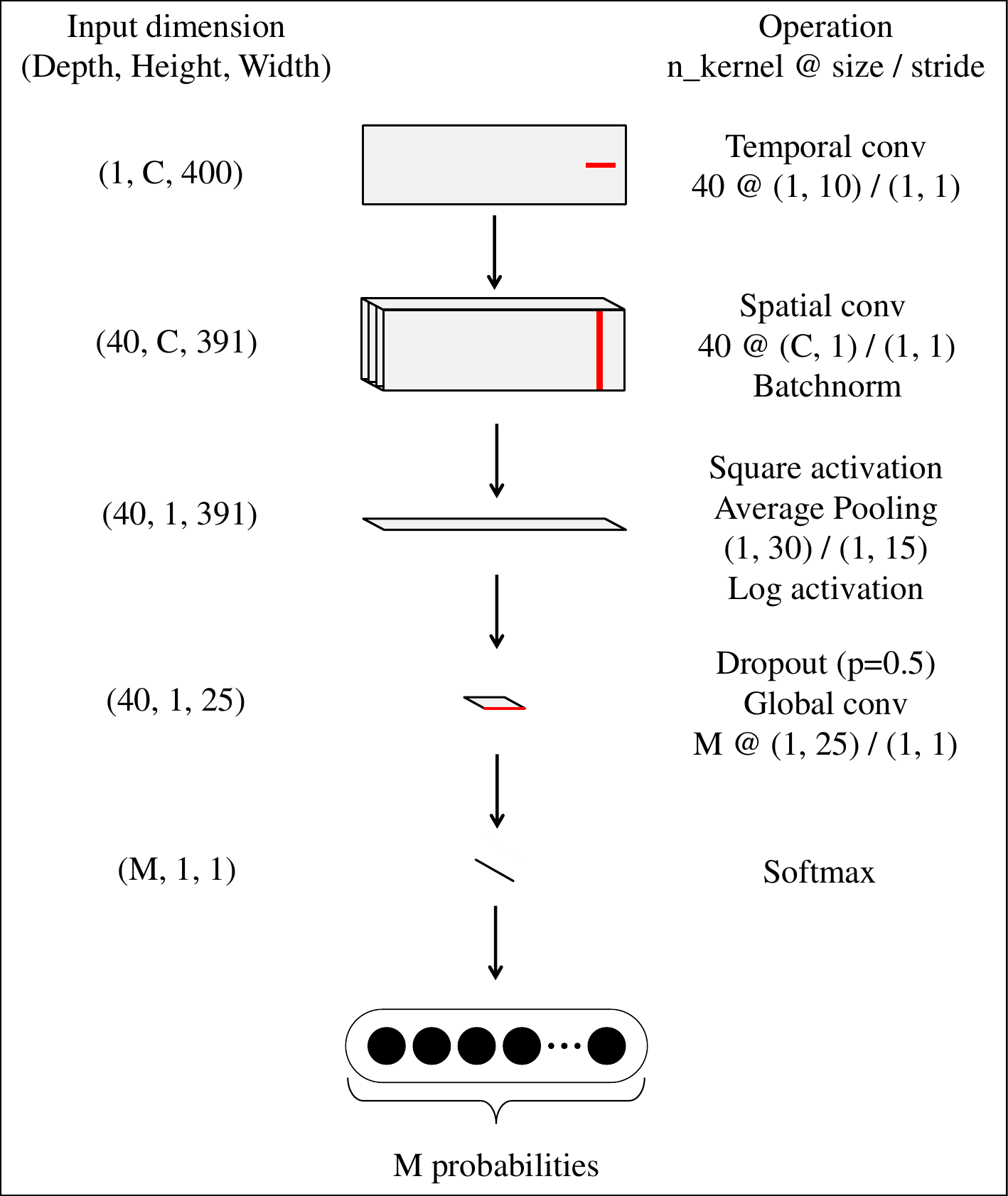}
	\caption{Architecture of the CNN classifier, $C$ and $M$ denotes the number of channels and classes respectively. Convolution kernels are presented in red.}
	\label{fig.cnn}
\end{figure}

A convolutional neural network (CNN) with shallow architecture \cite{schirrmeister2017deep} is employed as the classifier to evaluate the efficacy of the proposed channel selection method, as it shows efficacy on MI classification in the literature and our preliminary experiments.
The architecture of the CNN classifier is shown in Fig. \ref{fig.cnn}.

The network takes a $C$-channel EEG trial of length $T$ as input, which can be represented by a matrix $E \in \mathbb{R}^{C \times T}$ (in this paper, EEG trials are cropped such that $T=400$).
Firstly, a convolutional layer of size (1, 10) with 40 kernels performs temporal convolution on the input, followed by a spatial convolutional layer with 40 kernels of size (C, 1) and a batch normalization layer, resulting a feature map $F_1 \in \mathbb{R}^{40 \times 1 \times 391}$.
Subsequently, a sequence of nonlinear operations, including a squaring activation function, an average pooling layer and a logarithmic activation function are imposed on $F_1$, producing a higher-level feature map $F_2 \in \mathbb{R}^{40 \times 1 \times 25}$.
Finally, a dropout layer with dropping probability $p=0.5$ is employed before a global convolution layer and a softmax layer transform $F_2$ into probabilistic predictions among $M$ classes.

\section{Experiments and Results}\label{sec.results}

In this section, the performance of the proposed channel selection method is evaluated on two publicly available motor imagery EEG datasets. 

\begin{figure}[h] 
	\centering 
	\subfloat[Dataset A]{
	  \includegraphics[width=0.46\linewidth]{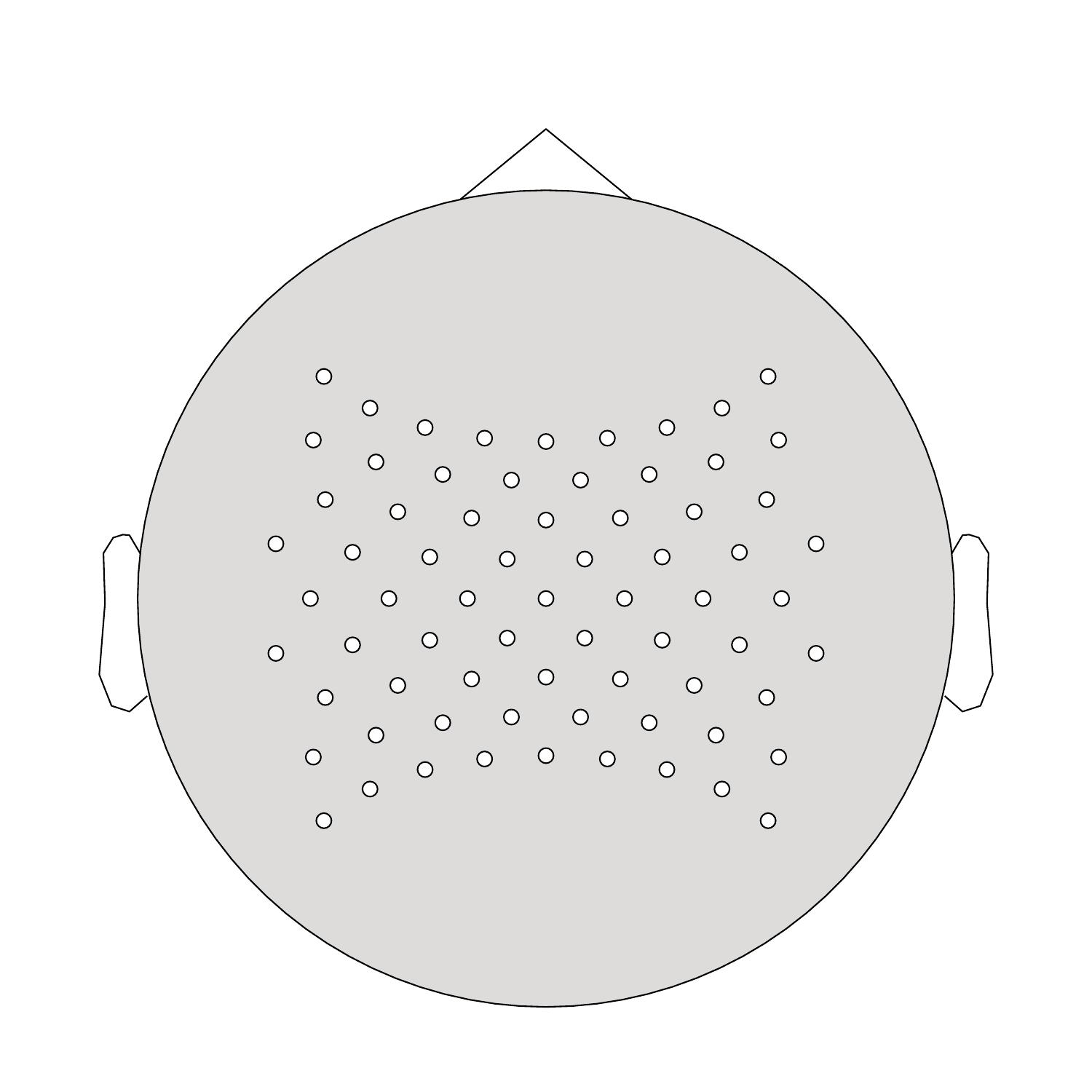}
	  \label{fig.4a_layout}
	  }
	\hfil
	\subfloat[Dataset B]{
	  \includegraphics[width=0.46\linewidth]{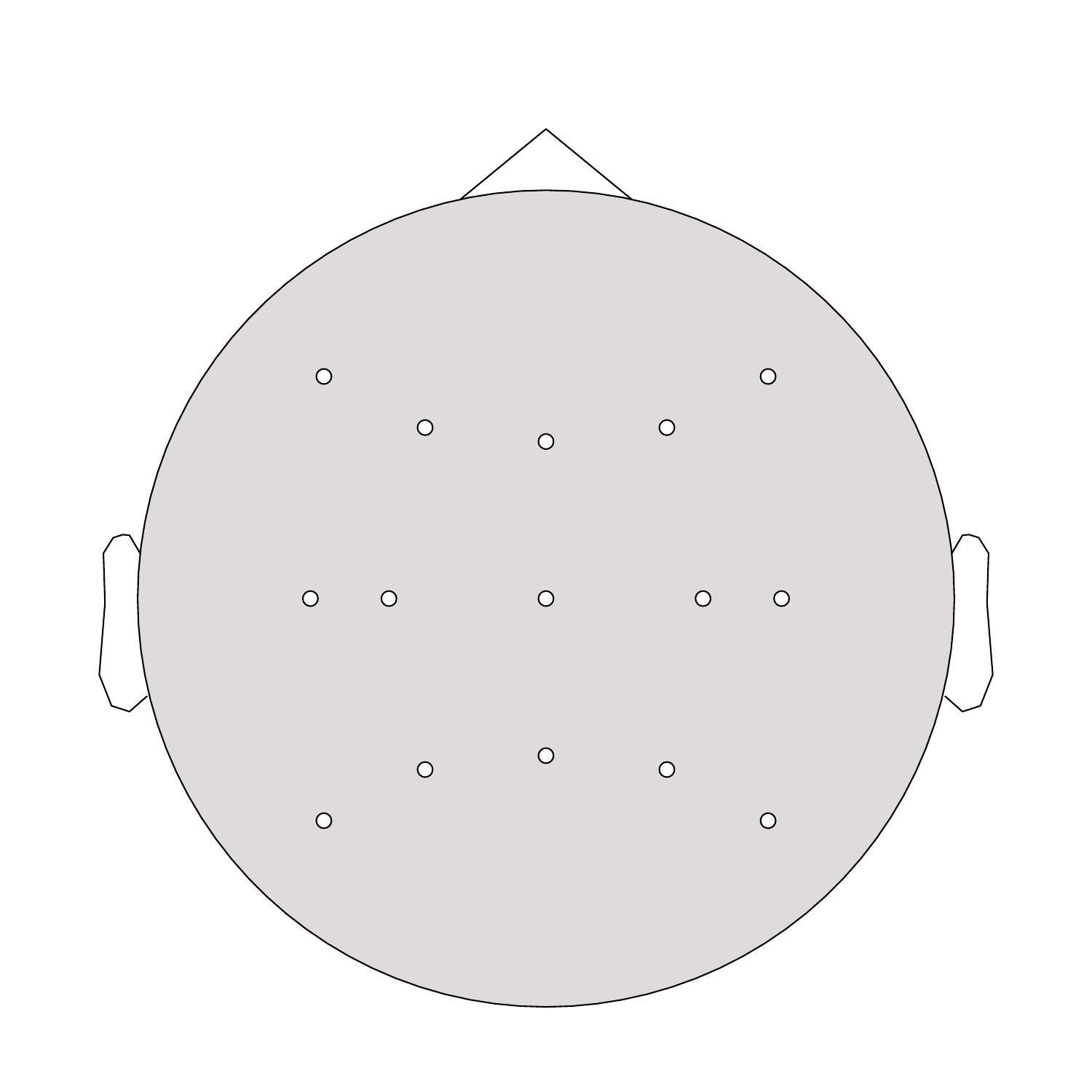}
	  \label{fig.cla_layout}
	  }
	\caption{Electrode layouts of two datasets, corresponding to the all-channel setups in the experiments. Positions of electrodes are marked by white dots. Electrodes between and including F (frontal) column and P (parietal) column in the international 10-20 system are preselected for evaluation, resulting in 71-channel and 15-channel layouts for datasets A and B respectively.} \label{fig.layout}
\end{figure}

\subsection{EEG Dataset Description}
\subsubsection{Dataset A (BCI Competition III Dataset IVa)}
This dataset contains EEG signals from five subjects performing right hand and right foot motor imagery tasks.
For each subject, 280 EEG trials are recorded at a sample rate of 1kHz from 118 electrodes positioned according to the extended international 10-20 system.
The experiment was conducted under a classical cue-based paradigm.
Each trial starts with a visual cue indicating one of three types of motor imagery tasks (namely left hand, right hand and right foot).
Subjects were asked to perform the correspondent MI task for 3.5 seconds before a period of random length between 1.75 to 2.25 seconds for relaxation.
Only the trials of `right hand' and `right foot' tasks are provided in the dataset.
In this paper, we randomly selected 20\% of the trials of each subject as the test set to evaluate the efficacy of the proposed method.
Besides, 10\% of the remaining trials are randomly selected and designated as the validation set in the training procedure.

\subsubsection{Dataset B (CLA Dataset)}
This dataset is a subset of a large EEG dataset \cite{kaya2018large}, in which all experiments were approved by the Ethics Committees of Toros University and Mersin University.
Dataset B contains MI trials in the classical (CLA) left/right hand MI paradigm.
EEG signals are recorded at a sampling frequency of 200 Hz from 19 electrodes in the standard international 10-20 system, along with two ground electrodes (A1 \& A2) and a synchronization channel which does not contain actual EEG data.
At the beginning of each trial, a visual signal indicating left hand, right hand or passive response is shown on a screen for 1 s, and the subjects need to implement motor imagery in response.
In this paper, trials of left hand and right hand MI tasks from subjects with three sessions of EEG record (namely subjects B, C, E, F and J) are employed.
Each session contains 616 to 636 trials, with subject F as an exception whose first two sessions contain 960 and 652 trials respectively.
For each subject, the last session is used as test set, 10\% of trials in the other two sessions are randomly chosen as validation set, and the trials remained are used as training set.

Based on the existing neurophysiological researches \cite{mcfarland2000mu,pfurtscheller2006mu}, the electrodes on the scalp near sensorimotor cortex are preselected for analysis.
The layouts of EEG channels used in this article are presented in Fig. \ref{fig.layout}.

\begin{figure}[!t]
	\centering
	\subfloat[Dataset A]{
	  \includegraphics[width=3.2in]{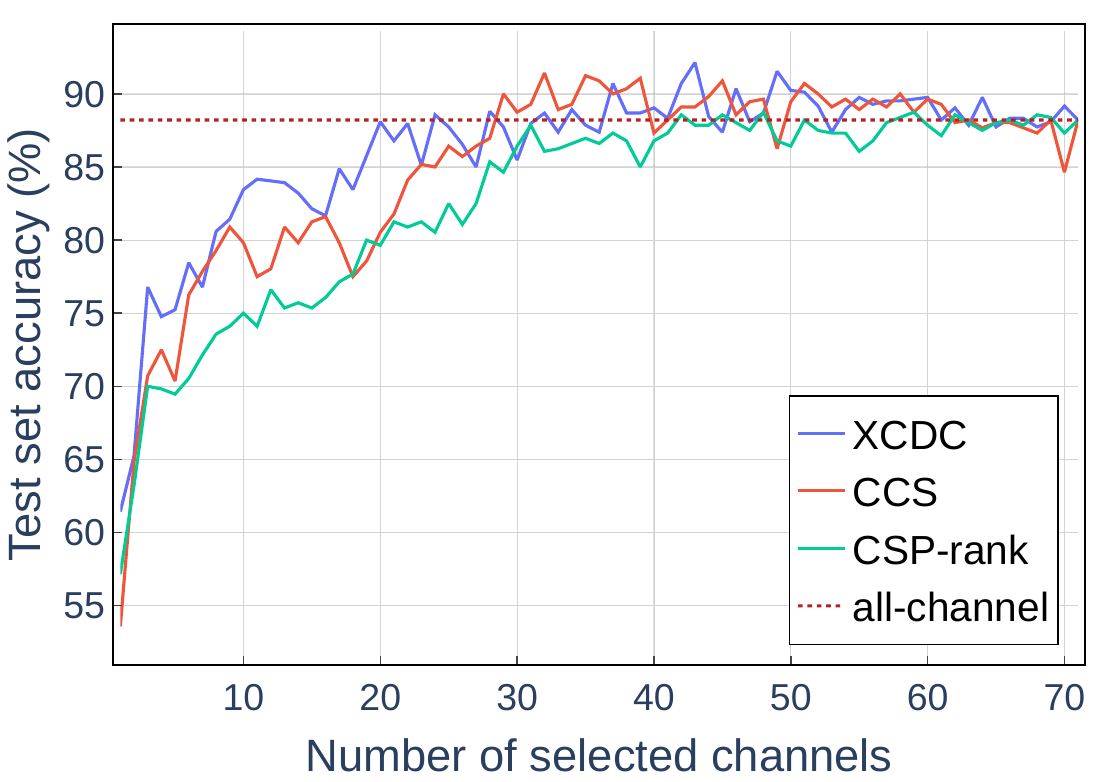}
	  \label{fig.4a_acc_comparison}
	  }
	\hfil
	\subfloat[Dataset B]{
	  \includegraphics[width=3.2in]{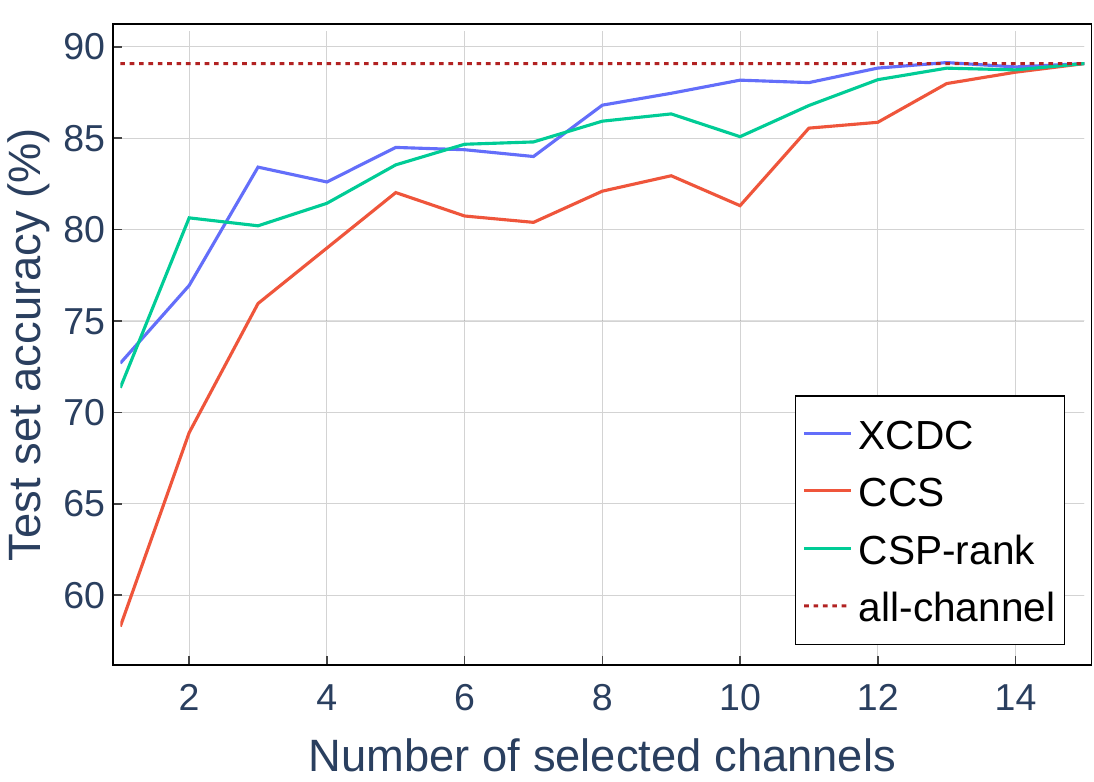}
	  \label{fig.cla_acc_comparison}
	  }
	\caption{Comparison between the proposed channel selection method with CCS and CSP-rank on average test set accuracy over all subjects from both datasets. Dotted lines indicate the accuracy of all-channel setups.}
	\label{fig.acc_comparison}
\end{figure}

\begin{figure*}[htbp]
	\centering
	\includegraphics[width=0.95\linewidth]{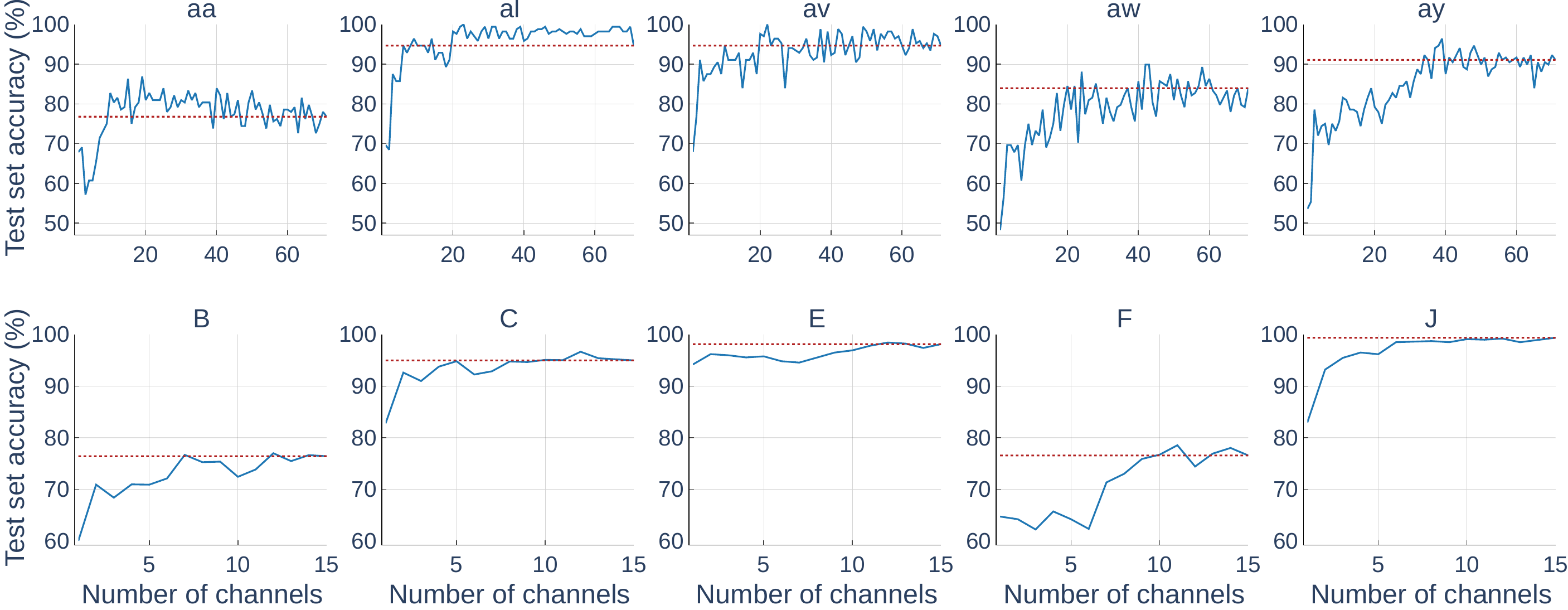}
	\caption{Performance of XCDC for all subjects in both datasets.
	The upper row depicts accuracies of subjects in dataset A, and the bottom row presents accuracies of subjects in dataset B.
	Dotted lines show the accuracy of the all-channel setup, corresponding to an accuracy decrease constraint of 0\%.
	}
	\label{fig.acc_subjects}
\end{figure*}

\subsection{Data Processing}
Firstly, the following steps are taken to pre-process the signals before channel selection and classification:
\begin{enumerate}
    \item Raw EEG signals are filtered using a 2nd-order Butterworth filter with a pass-band range of 0.1-30 Hz and then downsampled to 100 Hz.
    \item Signals are cropped with a time window of 0-4 s relative to the visual cue onset, resulting a length of 400 samples per trial.
    \item Z-score normalization are employed with respect to each channel, making the average value of signals form each channel to be 0 and standard deviation to be 1.
\end{enumerate}
After preprocessing, channels are ranked according to the discriminant criterion $D$ presented in (\ref{eqn.6}).
Given a designated number $k$, signals from the top-$k$ channels are selected to form the input of the CNN classifier.
In order to alleviate the effect of orders of channels, rows of the input matrices are spatially arranged (left to right, frontal to parietal) according to the electrode layouts.
Subsequently, the CNN classifier is trained on the training set for 500 epochs with the batch size of 120.
In the training procedure, Adam \cite{kingma2014adam} is used as the optimizer, whose learning rate and weight decay parameters are determined before the experiments using 5-fold cross validation and Bayesian Optimization \cite{snoek2012practical} with signals from all the channels as input.
Finally, the test set classification accuracy of the trained classifier is used as the metric to evaluate the proposed method. 
In the experiments, 10-fold cross validation is employed on the training set to determine the subject-specific weighting parameter $\lambda$ in (\ref{eqn.6}), using signals from the top-3 channels as input.

\subsection{Results}

\subsubsection{Channel Selection}
Test set classification accuracy using the top-$k$ channels is evaluated with $k$ ranging from 1 to all the channels (71 and 15 channels for datasets A and B respectively).
For the purpose of comparison, Correlation-based Channel Selection (CCS) \cite{jin2019correlation} and CSP-rank \cite{tam2011performance} are also implemented and evaluated using the same classifier. 
In the experiments, hyper parameters of the CNN classifier are identical for all the comparing methods.
Furthermore, we also analyzed the number of channels that can be removed while satisfying certain constraints of accuracy.

\begin{figure*}[!t]
	\centering
	\subfloat[Dataset A, right hand vs. right foot]{
	  \includegraphics[width=0.98\linewidth]{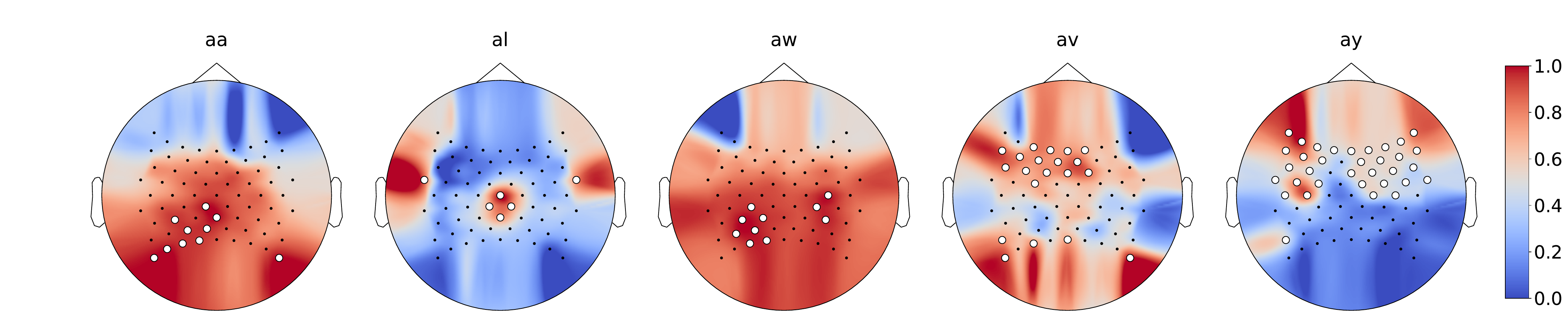}
	  \label{fig.4a_viz}
	  }
	\hfil
	\subfloat[Dataset B, left hand vs. right hand]{
	  \includegraphics[width=0.98\linewidth]{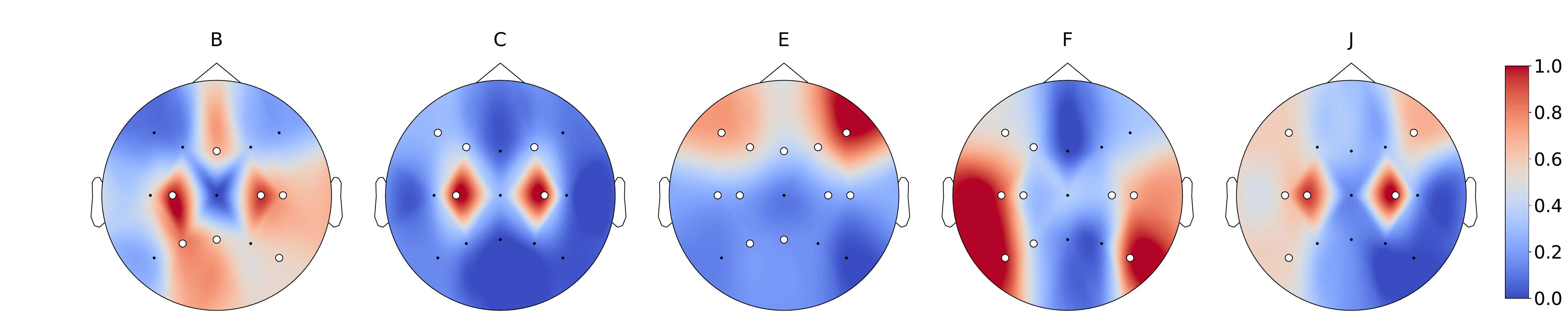}
	  \label{fig.cla_viz}
	  }
	\caption{Visualization of normalized discriminant score $D$ of all subjects. 
	White markers indicate the minimal channel subsets to achieve accuracies within a decrease under 1\% compared to all-channel setups. 
	The position of other channels are marked by black dots, and values between channels are obtained through bilinear interpolation.
	For dataset A in which subjects perform right hand and right foot motor imagery tasks, most of the important electrodes are distributed near the center region or contralateral to the performing side for subjects aa, al, aw and av. 
	For dataset B in which subjects implement left/right hand motor imagery, task relevant regions distinguished by the proposed method show a symmetric pattern.}
	\label{fig.disc_score_viz}
\end{figure*}

Fig. \ref{fig.acc_comparison} compares the average accuracy of different  methods against the number of channels.
Fig. \ref{fig.4a_acc_comparison} shows that for dataset A, XCDC still maintains comparable performance compared to all-channel accuracy when the number of channels is reduced from 71 to 20, and outperforms both CCS and CSP-rank when the channel number is small.
Fig. \ref{fig.cla_acc_comparison} shows that XCDC globally outperforms CCS, and gives higher classification accuracy than CSP-rank with few exceptions at 2, 6, and 7 channels.
It can be seen by comparing Fig. \ref{fig.4a_acc_comparison} and Fig. \ref{fig.cla_acc_comparison} that, for dataset A, removing less important channels yields better performances for both CCS and XCDC.
Whereas the accuracy of dataset B keeps a downward trend with the channel removal.
This may be due to the noise and redundancy brought by a large amount of electrodes in dataset A.

Fig. \ref{fig.acc_subjects} shows subject-specific accuracy against channel numbers.
The top row depicts accuracies on dataset A and the accuracies on dataset B are illustrated at the bottom row.
As is clearly shown in the figure, for subjects aa, al and av in dataset A, the number of channels can be reduced from 71 to no more than 10 without significant loss in accuracy.
It is also notable that for subject E in dataset B, using only one channel is sufficient to have an accuracy above 94\%.
By taking a closer look at Fig. \ref{fig.acc_subjects}, we can see that, despite the fluctuations, with the channels with lowest $D$ being removed (corresponding to the rightmost sections of the curves), the accuracies maintain unharmed or show an overall tendency of increase, but removing more channels results in a decrease of classification accuracy.
This implies that channels with low discriminant scores are redundant or irrelevant to the mental task, and the relevant channels are distinguished and given higher rankings by XCDC.

\begin{table*}[htbp]
	\centering
    \caption{Number of Channels in the Minimal Subset Under Accuracy Decrease Constraints}
    \begin{tabular}{cccccccccc}
    \toprule
    ~       & \multicolumn{9}{c}{Accuracy Decrease Constraint}                                                                                   \\ \cmidrule{2-10} 
    Subjects & \multicolumn{3}{c}{5\%}    & \multicolumn{3}{c}{1\%}    & \multicolumn{3}{c}{0\%}    \\ 
    \cmidrule(r){2-4} \cmidrule(r){5-7} \cmidrule{8-10} 
    ~       &  XCDC     & CCS  & CSP-rank & XCDC     & CCS  & CSP-rank & XCDC     & CCS  & CSP-rank \\ \midrule
    aa              & 8        & 6    & 6        & 10       & 6    & 35       & 10       & 6    & 35       \\ 
    al              & 6        & 3    & 3        & 6        & 3    & 7        & 9        & 3    & 7        \\ 
    av             & 3        & 7    & 19       & 10       & 9    & 28       & 20       & 9    & 30       \\ 
    aw             & 17       & 13   & 13       & 20       & 60   & 47       & 20       & 60   & 47       \\ 
    ay             & 32       & 28   & 29       & 34       & 29   & 30       & 34       & 29   & 30       \\ \midrule
    mean           & 13.2     & \textbf{11.4} & 14       & \textbf{16}     & 21.4 & 29.4     & \textbf{18.6}     & 21.4 & 29.8     \\ \midrule
    B               & 7        & 13   & 13       & 7        & 15   & 13       & 7        & 15   & 15       \\ 
    C               & 2        & 11   & 2        & 5        & 13   & 6        & 10       & 13   & 12       \\ 
    E               & 1        & 4    & 7        & 11       & 10   & 13       & 12       & 14   & 15       \\ 
    F               & 8        & 11   & 7        & 9        & 12   & 8        & 10       & 14   & 9        \\ 
    J               & 3        & 5    & 2        & 6        & 12   & 8        & 15       & 15   & 11       \\ \midrule
    mean          & \textbf{4.2}      & 8.8  & 6.2      & \textbf{7.6}      & 12.4 & 9.6      & \textbf{10.8}     & 14.2 & 12.4     \\
    \bottomrule
    \end{tabular}\label{table.1}
\end{table*}

We further investigated how many channels can be removed while meeting certain requirements of accuracy.
Given a reference accuracy $a_r$ and a constraint of relative accuracy decrease $d$, a set of channel setups within the constraint is obtained:
\begin{equation}
	A = \{k|a_k \geq a_r(1-d)\}, k=1, 2, \dots, C
\end{equation}
where $C$ denotes the channel number of the all-channel setup and $a_k$ is the classification accuracy using the top-$k$ channels in the ranking.
Then we look for the setup with fewest channels in $A$, termed minimal subset:
\begin{equation}
	k_m = \mathbf{min}(k|k \in A)
\end{equation}
For each subject, we choose the all-channel accuracy as the reference accuracy $a_r$, and investigated $k_m$ for XCDC, CCS and CSP-rank with the constraint $d$ set to 5\%, 1\% and 0.
The results are presented in Table \ref{table.1}, the best results amongst the compared algorithms are highlighted in boldface.
The table shows that for both datasets, the proposed method uses lower average number of channels than the comparing methods in most cases.
More specifically, using XCDC, the number of channels can be reduced to an average of 16 and 7.6 for datasets A and B respectively while ensuring a relative accuracy decrease under 1\%, which is better than the results for CCS and CSP-rank.
Note that for subjects aa, al and aw, the number of channels can be reduced from 71 to 6 or 10 using XCDC when $d=1\%$, whereas more than half of the electrodes can also be removed for subjects av and ay under the same constraint.

\subsubsection{Visualization}
Visualization results show that XCDC is neurophysiologically meaningful.
The discriminant scores $D$ of each subject are normalized to a range of [0, 1] and visualized in Fig. \ref{fig.disc_score_viz}, the minimum subset with $d=1\%$ are marked by white dots.
Values at the locations without channels are obtained through bilinear interpolation for the convenience of presentation.
We can see that, for dataset A in which subjects perform right hand versus right foot motor imagery tasks, the task related electrodes of subjects aa, al, aw and av are mainly distributed near the center region of the cortex or contralateral to the performing side.
Besides, it is shown by Fig. \ref{fig.cla_viz} that, with the subjects in dataset B implementing left/right hand MI, task relevant regions show a symmetric pattern.
Note that for subjects B, C and J, channels C3 and C4 are distinguished by XCDC as the most important channels, which is consistent with prior neurophysiological findings.

\section{Discussion}\label{sec.discussions}

\subsection{Channel Ranking and Channel Selection}
Finding the optimal subset of channels by evaluating every possible combination is infeasible.
Many channel selection techniques, the one proposed in this article included, narrow down the range of choices by performing rather a channel ranking than a channel selection. 
These methods produce a sorted list of channels, either by ranking according to a certain numerical criterion, or by removing channels sequentially.
Some methods, CSP with regularization terms for instance, give a specific subset of channels, but the number of channels in the result differs while the regularization parameter changes, and the users still need to assess the performance of different combinations before finally choosing the optimal channel setup.
However, there are still two problems remain to be addressed for this type of methods: 
\begin{inparaenum}[1)]
    \item how to make comparisons between channel ranking methods, and 
    \item how to choose a subset given the result from these methods.
\end{inparaenum}

In practice, channel selection is more of a trade-off between performance and simplicity.
Some articles compare channel selection methods by the highest accuracy among all the subsets, whereas some other researches make comparison by classification accuracy on several channel numbers.
However, we can see from Fig. \ref{fig.acc_subjects} as well as many other researches that, the classification accuracy usually fluctuates as the number of channels changes.
This makes the above methods less practical for both comparison and applications, since comparable performance can possibly be achieved using much fewer channels.
Take subject B in Fig. \ref{fig.acc_subjects} for instance, although the classification accuracy reached its peak at 76.99\% using 12 channels, the 7-channel setup gives a comparable result (77.68\%) within a gap of 0.31\%.

This article provides an instructive and practical way to solve the two aforementioned problems: in the calibration phase, designate a reference accuracy and a tolerance rate of accuracy deterioration, then select the subset meeting the requirement with fewest channels.
In this article, the all-channel accuracy is chosen as reference, and the tolerance rates of 5\%, 1\% and 0 are evaluated.
Nonetheless, these two parameters can be arbitrarily designated in applications according to actual requirements.
When the highest accuracy is chosen for reference and the tolerance rate is set to 0, this method degenerates to selecting the setup with maximum accuracy.

\subsection{Computational Complexity}
The complexity of XCDC is quadratic.
Deriving the discriminant score $D$ of a channel involves the computation of cross-correlation between every pair of trials, resulting in a quadratic complexity.
Besides, cross-correlation itself is a cumbersome operation with numerous multiplications and additions.
Therefore, if implemented directly by definition, the proposed method is unusable in terms of computation time in BCI system design.
For example, in the experiments, given 600 trials from a subject, it takes more than 8 seconds to obtain $D$ for one channel, making it impractical for deployment on BCI systems with large amount of channels, not to mention that the computation time scales quadratically with more trials.

However, this problem can be tackled by utilizing deep learning frameworks with GPU acceleration.
In the scope of deep learning techniques, the convolution operation with a stride of 1 in its one-dimensional case is identical to the correlating procedure in (\ref{eqn.1}).
Therefore, it is possible to implement the proposed method using deep learning libraries such as PyTorch, which can be significantly accelerated with a GPU backend\footnote{Experiments are conducted on a PC with an Intel Xeon E5-2670 CPU and an Nvidia GeForce GTX 980 GPU. GPU acceleration is supported by PyTorch library and CUDA toolkit.}.
In the experiments, the PyTorch implementation of XCDC drastically reduces the computation time, requiring less than 0.038 s under the aforementioned condition.
This makes the proposed method an efficient algorithm given that it contains no iterative optimization procedures.

\subsection{Limitations}
Despite the encouraging performance and efficiency, this article has important limitations.
First, experiments on other EEG datasets are required to further evaluate the performance of the proposed method.
Second, the way of choosing the weighting hyperparameter $\lambda$ is suboptimal.
One possible workaround for application is to determine the number of channels using a default setting of $\lambda$, and then optimize $\lambda$ specifically for the chosen number.

\section{Conclusion} \label{sec.conclusion}
This article proposed a novel channel selection method named XCDC for motor imagery EEG classification.
The proposed method selects the most discriminant channels based on cross-correlation between signals from different EEG trials.
Experiments were conducted on two EEG datasets to evaluate the performance.
Results show that XCDC successfully distinguishes the task-relevant channels and reduces the number of channels significantly without loss in classification accuracy, outperforming CCS and CSP-rank on both datasets.
Besides, visualization of the proposed method is consistent with prior neurophysiological findings.


\section*{References}
\bibliography{ref}
\bibliographystyle{unsrt}

\end{document}